\documentclass[a4paper,12pt]{article}
\usepackage[cp1251]{inputenc}
\usepackage{graphics}
\usepackage{amssymb,amsmath}
\usepackage[english]{babel}
\usepackage{indentfirst}
\begin{document}
\bigskip
\centerline {\bf Cycles of the magnetic activity of the Sun  and
solar-type stars and simulation of their fluxes}

\bigskip

\centerline {E.A. Bruevich $^{a}$ , I.K. Rozgacheva $^{b}$}
\bigskip

\centerline {Sternberg Astronomical Institute, Moscow, Russia}\
\centerline  {Moscow State Pedagogical University, Russia}\
\centerline {E-mail: $^a${red-field@yandex.ru},
$^b${rozgacheva@yandex.ru}}\

\bigskip
{\bf Abstract.}
      The application of the Wavelet analysis and Fourier analysis to the dataset of variations
      of radiation fluxes of solar-like stars and the Sun is examined.
      In case of the Sun the wavelet-analysis helped us to see a set of values
      of periods of cycles besides "11-year" cycle: the long-duration cycles
      of 22-year, 40-50 year and 100-120 year and short-duration cycles of 2-3,5 years and 1,3-year.
We present a method of the chromospheric flux simulation using the
13 late-type stars, which have well-determined cyclic flux
variations similar to the 11-year solar activity cycle. Our flux
prediction is based on the chromospheric calcium emission time
series measurements from the Mount Wilson Observatory and comparable
solar dataset. We show that solar three - component modeling well
explains the stellar chromospheric observations.

\bigskip

\bigskip
\vskip12pt {\bf1 . Introduction. The solar-type magnetic activity
among the stars } \vskip12pt

    Magnetic activity of the Sun is called the complex of composite
electromagnetic and hydrodynamic processes in the solar atmosphere.
They create a local active area: plages and spots in the
photosphere, calcium flocculi in the chromosphere and prominences in
the corona of the Sun. The analysis of active regions is necessary
to study the magnetic field of the Sun and the physics of magnetic
activity. This task is of fundamental importance for astrophysics of
the Sun and the stars. Its applied meaning is connected with the
influence of solar flares on the Earth's magnetic field and the need
to solar activity variations and flares forecasts.

It is difficult to predict the the evolution of each active region
in details in present time. However, it has long been established
that the total change of the active areas is cyclical. The cyclical
nature of solar activity  allows you to predict the state of the
global activity of the Sun.

The duration of the "eleven-year cycle of solar activity ranged from
7 to 17 years according to 160 years of direct solar observations.

For the first time the quasi-biennial variations of solar activity
have been described in the work of the (Vitinsky et al. 1986).

In the recent studies on the subject of quasi-biennial variations of
solar radiation (Ivanov-Kholodnyj \& Chertoprud 2008; Bruevich \&
Ivanov-Kholodnyj 2011) the importance of  this problem study  were
emphasized. It turned out that quasi-biennial solar cycles are
closely associated with various quasi-biennial processes on the
Earth, in particular with quasi-biennial  variations of the velocity
of the Earth rotation and speed of the stratospheric wind.

The main methods of quasi-biennial solar cycles and the results of
their study were more fully described in the monograph (Rivin 1989).

Application of modern mathematical methods (modification of the
method of main components, the "Singular Spectrum Analysis") for the
treatment of long time series of 3032 the monthly averaged values of
Wolf numbers gave an opportunity to draw some conclusions about the
duration of the cycle and the form of individual quasi-biennial
oscillations depending on the duration and power of eleven-year
cycle (Khramova et al. 2002).

It was discovered the activity with the cyclicity period of 1,3
years during the last eight of the 11-year cycle. It is best
manifested in the phase of maximum of the 11-year cycle and in the
early phase of its decline. The axis of the dipole large-scale field
at this time is located in the plane of the solar equator (Livshits
\& Obrydko 2006). These facts show on the independent existence of
large-scale magnetic field and its influence on the processes of
local magnetic activity.

A comparison of magnetic activity of solar-type stars of different
age  allows us to check the basic representations of the internal
structure and evolution of convective shells of these stars.

Among the hundreds of thousands of stars in the vicinity of the Sun
only few thousands fall under the definition of solar-type stars.
The relative paucity of such stars is a consequence of their low
luminosity. This prevents their detection and observation at a
considerable distance from the Sun. Photometric observations of
stars with active atmospheres are regularly held in the optical
range since the mid of XX century. These observations include the
measurement of their radiation in different ranges of the
electromagnetic spectrum during the long intervals of time.

In the present work we study the low-amplitude  cyclical variability
of solar and star's fluxes, as well as the duration of cycles and
their dependence on the physical parameters of the Sun and
solar-type stars.
\bigskip
\bigskip
\bigskip
\vskip12pt {\bf1.1  "HK-project" of Mount Wilson observatory }
\vskip12pt

One of the first and still the most outstanding program of
observations of solar-type stars is the observation programme
"HK-project" of Mount Wilson observatory.

Implementation of this project has led to the discovery of
"11-cycles" activity in solar-type stars. This observation program
lasts for more than 40 years.

First O. Wilson began this program in 1965. He attached great
importance to the long-standing systematic observations of cycles in
the stars (Baliunas et al. 1995; Lockwood et al. 2007).

For this project the stars were carefully chosen according to those
physical parameters, which were  most close to the Sun: cold, single
stars-dwarfs, belonging to the Main sequence. Close binary systems
are excluded. All the stars of "HK-project" almost evenly are
distributed on the celestial sphere.

In the framework of the "HK-project" information about the
chromospheric fluxes variations of the Sun was obtained from
ground-based observations of the global index of solar activity (the
10,7 cm radio flux) - F10,7, and these observations were
subsequently adapted to the same values of star's fluxes in the
chromospheric lines of $H$ (396.8nm) and $K$ (393.4nm) $Ca II$.

Thus, in our work we use for stars and the Sun S-index (or
$S_{CaII}$) - the relationship of radiation fluxes in the centers of
emission lines $H$ and $K$ (396,8 nm and 393,4 nm) to radiation
fluxes in the near-continuum (400,1 and 390,1 nm). This index is a
sensitive indicator of the chromospheric activity of the Sun and the
stars.

The first results concerning the observations of 91 stars, were
published in the (Wilson 1978). Since 1974 the head of the project
became A. Vaughan. He constructed a "HK-spectrometer" of the next
generation with the use of more modern technologies and continued
observation of several hundred stars. At the same time observations
of the same stars with help of  the 60-inch telescope took place to
determine their periods of rotation.

The results of the joint observations of the radiation fluxes and
periods of rotation gave the opportunity for the first time in
stellar astrophysics to detect the rotational modulation of the
observed fluxes (Noyes et al. 1984). This meant that on the surface
of the star there are inhomogeneities those were living and evolving
in several periods of rotation of the stars around its axis. In
addition, the evolution of the periods of rotation of the stars in
time clearly pointed to the fact of existence of the star
differential rotations similar to the Sun differential rotations.

Radiation fluxes in the lines of $H$ and $K$ $Ca II$  from the Sun
are formed on the upper levels of the solar of chromosphere and they
are good indicators of active regions - areas with high magnetic
activity.

Because of the remoteness of the stars from us we can't distinguish
the different active regions on their disks. However, the study of
the radiation fluxes in the lines of $H$ and $K$ $Ca II$.,
normalized on a nearby continuum, gives us the indirect information
about the numbers and sizes of the active regions in the atmospheres
of the stars.

To study the cycles of magnetic activity of the atmospheres of the
stars we use new dataset of simultaneous observations of variations
of the photospheric and chromospheric radiation fluxes of the Sun
and of 33 stars of "HK-project".

These data were obtained during the last 20 years in Lowell
observatory (photometric observations) and during 40 years of
observation in chromospheric lines of 111 stars in Smithsonian
observatory of Stanford University (Baliunas et al. 1995; Radick et
al. 1998; Lockwood et al. 2007).

The authors of "HK-project" with the help of frequency analysis of
the 40-year observations have discovered (Baliunas et al. 1995;
Lockwood et al. 2007) the periods of 11-year cyclic activity vary
little in size for the same star. So they have determined the
resistant variations of chromospheric activity. The durations of
cycles vary from 7 to 20 years for different stars. The stars with
cycles represent about 30\% of the total number of "HK-project"
stars. The determination of quasi-biennial activity of the cyclical
nature of the stars by these authors did not take place.

 The heterogeneities on the disk of the star, which are responsible for the atmospheric activity
 outside the flares are spots and plages in the photosphere, flocculi in the chromosphere,
 prominences and coronal mass ejections in the corona (by analogy with the
 thoroughly studied formations in the Sun).

 For the duration of the existence
 these active regions vary greatly in time - they can be observed from several hours to several months.
 The contribution from the active regions should be taken into account in the analysis
 of observational "HK-project" data, both as in the lines of H and K, as
 in photospheric fluxes from the entire disk in broadband filters of
 photometric system of Lowell observatory, close to the standard UBV - system.

\vskip12pt {\bf1.2  Cycles of magnetic activity of the Sun}
\vskip12pt

Variations of different indices of solar activity characterize the
variation of the radiation fluxes of the solar atmosphere at
different altitudes. In the works (Kane 2002; Li \& Sofia 2001) was
found a strong correlation of the index of EUV (short-wave
ultraviolet radiation, satellite observations of AE-E (1977-1980
years), Pioneer Venus (1979-1992 years) and SEM/SOHO (1996-2001
years) with the fluxes in the hydrogen lines , radio fluxes F10,7
and radiation fluxes in  the lines of $MgII$ (Beer 2000).

Also, all the indices of solar activity reasonably well correlated
with Wolf numbers. In Fig. 1 the annual average variations of Wolf
numbers from 1700 to 2005 years are shown. The values of W in the
period from 1850 to 2005 years were received as the result of a
direct observations of the solar activity with ground-based
observatories, and from 1750 to 1850 these values were obtained as a
result of indirect estimations.

\begin{figure}[h!]
 \centerline{\includegraphics{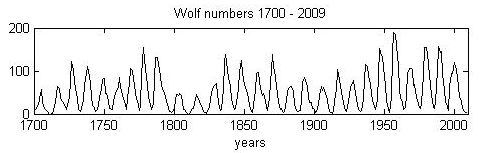}}
\caption{The time series of annual average of Wolf numbers from 1700
to 2005 years. Used data of the National Geophysical Data Center
Solar and Terrestrial Physics.}\label{Fi:Fig1}
\end{figure}

The result of wavelet - analysis (Daubechies wavelet) of series of
observations of average annual Wolf numbers   (Fig. 2) in the form
of many of isolines is shown in Fig. 2. For each isolines the value
of the wavelet-coefficients are of the same. The isolines specify
the maximum values of wavelet-coefficients, which corresponds to the
most likely value of the period of the cycle. There are three
well-defined cycles of activity:

 - the main cycle of activity is approximately equal to
a 10 - 11 years;

 - 40-50- year cyclicity;

  - 100 to 120-year-old (ancient) cyclicity.

\begin{figure}[h!]
 \centerline{\includegraphics{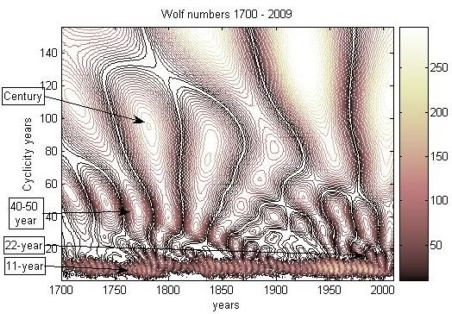}}
\caption{Wavelet-analysis (Daubechies wavelet) of time series of
annual averages of Wolf numbers. The ordinate axis is the duration
of the cycles (Cyclicity, years), the abscissa axis is the time
(years)}\label{Fi:Fig2}
\end{figure}

To identify a more short cycles we use the more accurate
measurements of Wolf numbers according to observations from 1950 to
2011 years. In Fig. 3 presents monthly averages of Wolf numbers.
These values exclude almost rotational modulation of the Sun's
observations, because the rotation period of the Sun on the "active
latitudes" is about 27 days, close to the interval on which
performed averaging.

\begin{figure}[h!]
 \centerline{\includegraphics{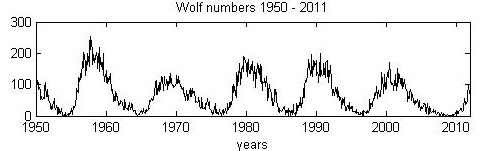}}
\caption{The time series of monthly mean values of Wolf numbers from
1950 to 2011 years, the  cycles 19 - 23. Used data of the National
Geophysical Data Center Solar and Terrestrial
Physics.}\label{Fi:Fig3}
\end{figure}

Fourier-analysis of time series of monthly average Wolf numbers can
not detect quasi-biennial cycles because of the considerable
variation in the duration of quasi-biennial cycles during the
"eleven-year" cycle. Wavelet-analysis gives us that opportunity.

Wavelet-analysis of a number of observations average monthly Wolf
numbers with the help of the wavelet Morley (Fig. 4)  shows that
there is the main cycle of variations is about a 10 - 11 years. We
clearly see in Fig. 4 that there are also short cycles: - 5,5-
year-old cycles, - quasi-biennial cycles

The values of the duration and amplitude of these short cycles are
changing in different intervals of observations. In Fig. 4 shows
that the duration of the quasi-biennial cycle is reduced from 3,5
years at the beginning of each "eleven-year cycle" of up to 2 years
at the end of it.

\begin{figure}[h!]
 \centerline{\includegraphics{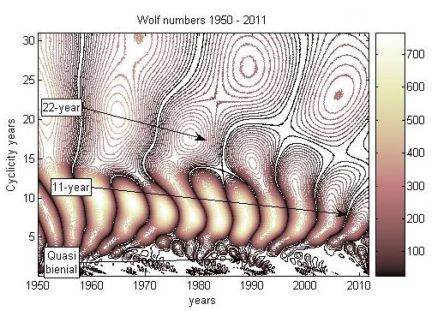}}
\caption{Wavelet-analysis (wavelet Morley) of the time series of
monthly average Wolf numbers. The ordinate axis is the duration of
the cycles (Cyclicity, years), the abscissa axis is the time
(years)}\label{Fi:Fig4}
\end{figure}

In Fig.5 and Fig.6 we show the results of our wavelet-analysis for
the numbers of the Wolf W and fluxes of radio emission F10,7 - the
most common indices of solar activity. On the axis of ordinates
postponed the duration of the quasi-biennial cycle, the x - axis the
time of observation in years.

\begin{figure}[h!]
 \centerline{\includegraphics{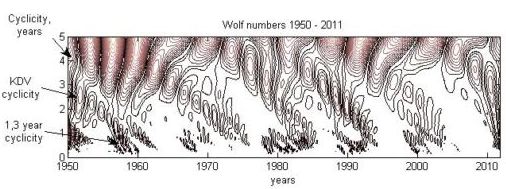}}
\caption{Wavelet-analysis of observations of Wolf numbers, data set
1950 - 2011 years. Used data of the National Geophysical Data Center
Solar and Terrestrial Physics.The ordinate axis is the duration of
the cycles (Cyclicity, years), the abscissa axis is the time
(years)}\label{Fi:Fig5}
\end{figure}

\begin{figure}[h!]
 \centerline{\includegraphics{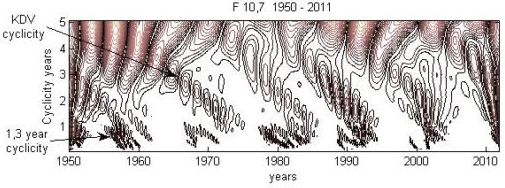}}
\caption{Wavelet-analysis of observations of fluxes of radio
emission $F_{10,7}$,  data set 1950 - 2011 years. Used data of the
National Geophysical Data Center Solar and Terrestrial Physics.The
ordinate axis is the duration of the cycles (Cyclicity, years), the
abscissa axis is the time (years)}\label{Fi:Fig6}
\end{figure}

The results of the wavelet-analysis, presented in Fig. 5 and Fig. 6
show that, along with  the quasi-biennial cyclicity of solar
radiation we can also see and cycles with a period of about 1,3,
found earlier in (Livshits \& Obrydko 2006)

\vskip12pt {\bf2. The cyclic activity of  solar-type stars on the
"11-year"  and quasi-biennial time scales}
\vskip12pt
In the last we
can see a large number of works, where low-amplitude variations of
the radiation of the stars with active atmospheres are investigated
with the help of modern approaches. In the work (Kolah \& Olach
2009) were analyzed variations of the radiation of EI Eri - one of
the bright solar-type stars. Data of observations were studied with
help of Fourier analysis and wavelet-analysis. Variations of the
fluxes of radiation of this star are quite stable. Study with help a
Fourier analysis or wavelet-analysis gives the similar results: the
cycle of variations of the fluxes of radiation is an average of 2,7
years.

\begin{table}
\caption{Results of our calculations of $T_{11}$ and $T_{2}$ values
and "HK-project" $T^{HK}_{11}$ calculations for 29 stars and the
Sun} \vskip12pt
\begin{tabular}{clclclclclclclcl}

\hline
    1&    2    &     3   &          4    &     5     &     6     &     7     &   8     \\ \hline
  No & Star    &Spectral &$P_{rot}, days$& $T_{eff}$, K &$T^{HK}_{11}$, &$T_{11}$,   &$T_{2}$ \\
     & on the HD& class   &(Soon et         &(Allen        &   years       &years       &years  \\
     & catalog &         &  al 1996)       & 1977)        &               &           &         \\ \hline
   1 &  Sun   & G2-G4   &    25           &     5780     &  10,0        &  10,7     &   2,7     \\
   2 &HD1835  & G2,5    &     8           &     5750     & 9,1          &   9,5     &   3,2     \\
   3 &HD3229  & F2      &     4           &     7000     & 4,1          &   -     &   -     \\
   4 &HD3651  & K0      &     45           &    4900     & 13,8          &   -     &   -     \\
   5 &HD4628  & K4      &    38,5         &     4500     & 8,37          &   -     &   -     \\
   6 &HD20630  & G5     &     9,24        &     5520     & 10,2          &   -     &   -     \\
   7 &HD26913 & G0      &     7,15        &     6030     & 7,8           &   -     &   -     \\
   8 &HD26965 & K1      &     43          &     4850     & 10,1          &   -     &   -     \\
   9 &HD32147 & K5      &     48          &     4130     &12,1           &   -     &   -     \\
   10&HD10476 & K1      &    35           &     5000     &  9,6         & 10       &    2,8    \\
   11&HD13421 & G0      &    17           &     5920     & -            & 10       &    -      \\
   12&HD18256 & F6      &    3            &     6450     & 6,8          & 6,7      &  3,2     \\
   13&HD25998  & F7     &    5            &     6320     &  -           &  7,1     &   -       \\
   14&HD35296 & F8      &   10            &     6200     &  -           & 10,8      &   -       \\
   15&HD39587 & G0      &    14           &     5920     &  -           & 10,5      &    -      \\
   16&HD75332 & F7      &    11           &     6320     &  -           &  9        &   2,4    \\
   17&HD76151 & G3      &    15           &     5700     &  -           &       -   &    $2,5^*$  \\
   18&HD76572 & F6      &    4            &     6450     &   7,1        & 8,5      &    -     \\
   19&HD78366 & G0      &    10         &     6030     &              &  10,2    &   -     \\
   20&HD81809  & G2     &   41           &      5780    & 8,2          & 8,5       &   2,0      \\
   21&HD82885 & G8      &   18           &      5490    & 7,9          & 8,6       &   -       \\
   22&HD100180& F7     &   14            &      6320    & 12           & 8         &    -       \\
   23&HD103095 & G8     &   31           &      5490    & 7,3          & 8         &    -       \\
   24&HD114710 & F9,5   &    12          &      6000    & 14,5        & 11,5       &   2      \\
   25&HD115383 & G0     &   12           &      5920    &  -           & 10,3      &    3,5      \\
   26&HD115404 & K1     &    18          &      5000    & 12,4         & 11,8      &   2,7     \\
   27&HD120136 & F7     &   4            &      6320    & 11,6         & 11,3       &    3,3      \\
   28&HD124570& F6      &    26          &      6450    & -            & -         &   2,7     \\
   29&HD129333 & G0     &   13           &      5920    & -            & 9         &    3,2      \\
   30&HD131156 & G2     &    6           &      5780    & -            & 8,5       &   3,8     \\

\hline
\end{tabular}
\end{table}

Table 1 and Table 2 present the results of our calculation of the
periods of variations of chromospheric radiation of the stars
(Bruevich \& Kononovich 2011). The method of fast Fourier transform
was used to determine periods of the cyclicity of stars.

\begin{table}
\caption{Table 1 - continued. Results of our calculations of
$T_{11}$ and $T_{2}$ values and "HK-project" $T^{HK}_{11}$
calculations for 23 stars}

\vskip12pt
\begin{tabular}{clclclclclclclcl}

\hline
    1&    2    &     3   &          4    &     5     &     6     &     7     &   8     \\ \hline
  No & Star    &Spectral &$P_{rot}, days$& $T_{eff}$, K &$T^{HK}_{11}$, &$T_{11}$,   &$T_{2}$ \\
     & on the HD& class   &(Soon et         &(Allen        &   years       &years       &years  \\
     & catalog &         &  al 1996)       & 1977)        &               &           &         \\ \hline
    31&HD143761 & G0     &   17            &     5920    & -            &   8,8     &    -      \\
   32&HD149661 & K2     &    21          &      4780    & 14,4         & 11,5      &   3,5     \\
   33&HD152391 & G7   &     11           &      5500    & 10,7         &  -        &     -     \\
   34&HD154417 & F8   &     7,8           &     6100    & 7,4          &   -        &    -     \\
   35 &HD155875 & K1    &     30           &      4850    & 5,7          &   -        &    -     \\
   36 &HD156026 & K5    &    21          &       4130    & -        &  11        &    -     \\
   37 &HD157856 & F6     &    4           &       6450   & -            & 10,9     &   2,6     \\
   38 &HD158614 & G9     &   34           &       5300   & -            & 12       &    2,6      \\
   39 &HD160346 & K3     &    37          &    4590      & 7           & 8,1        &   2,3     \\
   40 &HD166620 & K2      &   42           &       4780   & 15,8          & 13,7        &     -       \\
   41 &HD182572 & G8     &    41          &    5490      & -            & 10,5       &   3,1     \\
   42 &HD185144 & K0     &   27           &       5240   & -             & 8,5       &    2,6     \\
   43 &HD187681 & F8     &   10           &      6100    & 7,4            &  -        &    -     \\
   44 &HD188512 & G8     &   17           &       5490   & -            &  -        &    4,1     \\
   45 &HD190007 & K4     &    29          &      4500     & 10         & 11         &   2,5     \\
   46 &HD190406 & G1     &    14          &     5900     & 8        &   -        &     -     \\
   47 &HD201091 & K5     &   35           &      4410     &          & 13,1      &    3,6      \\
   48 &HD201092 & K7     &    38          &      4160     &          & 11,7      &   2,5     \\
   49 &HD203387 & G8     &   30            &       5490    & -            &  -        &    2,6      \\
   50 &HD206860 & G0     &    9          &     6300      & 6,2         & -        &  -     \\
   51 &HD216385 & F7     &    7           &       6320     & -           &  7        &   2,4    \\
   52 &HD219834 & K2     &    43          &      4780     & 10         & 11         &   2,5     \\
   53 &HD224930 & G3     &    33          &     5750    & 10,2        & -        &   -     \\
\hline
\end{tabular}

\end{table}

We used data of the observation of (Lockwood et al. 2007) for
frequency analysis of time series of S-index of the Sun and 33 of
the stars.

In the work (Bruevich and Kononovich 2011) the values of periods of
resistant "11-year"  cycles ($T_{11}$ - column 7 in Table 1 and
Table 2) and periods of quasi-biennial cycles ($T_2$ - column 8 in
Table 1 and Table 2) for the stars from column 2 of Table 1 and
Table 2 were determined. A dash in columns 7 and 8 means that the
cyclicity was not detected.

Table 1 and Table 2 contain the values of the "11-year" cycles
$T^{HK}_{11}$ (column 6), which were determined  by authors of the
"HK-project" during the conduct of primary spectral analysis of the
data (Baliunas et al. 1995).

Table 1 and Table 2 also show the spectral classes of stars (column
3), their periods of rotation $P_{rot}$ (column 4) and their
effective temperatures $T_{eff}$ (column 5).

The data analysis of Table 1 and Table 2 shows that among the stars
of the earlier spectral types, and accordingly, faster rotating, the
quasi-biennial cycles are found more confident. Note that these
cycles we found among the stars, as in the case of existing of
"11-year" cycles, so without them.
\vskip12pt
\vskip12pt {\bf3. The
cyclic variations of  fluxes of radiation from different layers of
the solar atmosphere}

\vskip12pt {\bf3.1 Indices of solar activity}
\vskip12pt

The first index, describing the solar activity, was the index of sun
spots. At the present time it is called the index of "Wolf numbers
". This index of solar activity has the most long-term history of
the direct observations. In addition, the rows of Wolf numbers were
restored since 1750 according to indirect data, see. Fig. 1. This
index characterizes the state of the photosphere of the Sun, so as
spots are such heterogeneities of that characterize the active
region at the level of photosphere.

Much more objective index of activity of photosphere is the index
Total Solar Irradiance (TSI). In addition to the fluxes of
photosphere's radiation (attributable to the visible range of the
spectrum) which give the main contribution into the TSI. The fluxes
of all the available spectral intervals from x-ray to infrared are
summarized in this index.

The most reliable are the data of TSI observations from 1978, with
help of the equipment installed on satellites HF, ACRIM-I,II and
VIRGO. Creation of the combined database of TSI observations Judith
Lean currently coordinates, author of a three-component model of EUV
(extreme ultraviolet) radiation of the Sun (Lean et al. 1986).

For monitoring of variations of chromosphere's  radiation of the Sun
on the satellites of the NOAA series from 1978 year to the present
time the measurements of the $Mg II$ (280 nm, core-to wing ratio)
index are made. This activity index is very similar to the S-index
of solar-type stars, observed in the "HK-project", with the
difference that in the case of the solar index $Mg II$ (280 nm)
observations are carried out in the range of radiation, out of reach
for the ground-based observatories.

Observations of solar radiation from the top of chromosphere and the
bottom of the corona the - index $F_{10,7}$ (radio flux at 10,7 cm
wavelength) are very important for solar activity study. This
activity index has also rather lengthy series of regular direct
observations with ground-based observatories (in Ottawa and other)
and correlates well with all the rest of the solar indices. Archive
data of index $F_{10,7}$  are detailed and accessible. In addition,
he is sensitive to changes in solar weather, the variation of this
index in the cycle of activity make up about 200\%, from ($75$ to
$240$)$\cdot 10^{-22}$ Watt/m2. Therefore, $F_{10,7}$  is used for
the prognosis and monitoring of the solar activity more often than
the other indices.

For observations of coronal activity of the Sun the index of x-ray
radiation of 0,1 - 0,8 nm (background of the radiation without
flares) is used. This activity index is very variable, even within a
small interval of time, because after even small flares the
background radiation remains elevated for some time. In a cycle of
activity  this index varies by two-three orders of magnitude from
$10^{-9}$ to $(1-5)$$\cdot10^{-7}$ Watt/m2.

Also there are another activity indexes associated with the flare
activity of the Sun, they are also well correlated with the main
solar indices. All activity indexes, characterizing fluxes of
radiation  from different layers of the solar atmosphere are
connected among themselves and are defined ultimately from the major
parameter of the activity - activity of the magnetic field.

\vskip12pt {\bf3.2 Empirical basis for the forecasts of the
amplitude of cyclic activity of stars} \vskip12pt For the Sun one of
the most important tasks is to predict the amplitude of cyclic
activity, affecting a number of earth processes. Contribution to the
low-amplitude variations of solar radiation occurs from the active
processes, simultaneously in the three layers of the atmosphere -
photosphere, chromosphere and the corona. It is known that different
indices of solar activity are rather closely interrelated. This fact
allows us to predict the variation of the solar radiation from
different layers, according to the observations of the change in the
index of activity from only one layer.

In the present work we use the ideology of the works (Lean et al.
1983; Lean 2001) for the prediction of "11-year" activity of the
stars. We present the full amplitude of the radiation heat flow in
the chromosphere's lines of stars in the form of the composition of
three main components of fluxes of radiation, two of which are the
constant and inconstant background with the grid of
supergranulation, the third is the contribution from the active
areas, which smoothly change during the cycle.

For the prediction of the amplitudes of variations of the fluxes of
radiation from the stars we chose the values of the fluxes in the
lines $H$ and $K$  $CaII$  of the "HK-project" stars  with the
well-identified "11-year-old" cycles (from 8 to 13 years). It turned
out that the cyclical nature of these fluxes is so similar to the
solar cycle, that for the forecast of amplitudes of fluxes of
radiation from these stars are quite applicable methods used in the
practice of solar forecasts (Borovik et al. 1997).

For the successful prediction of variations of the fluxes of
radiation from the stars, we should take into account the following
characteristics of their "11-year" activity cycles:

(a) The chromospheric emission lines $H$ and $K$, visible on the
background of wide and deep  profiles of absorption lines contain
information about the temperature of the atmosphere. It is
established, that the profiles of the lines differ markedly
depending on the photospheric substrate (is it the photosphere's
background, or the spot and the plage). The profiles of lines,
averaged over the disc of the Sun, depend on the phase of the solar
cycle;

(b) According to observations, sun spots are often collected in
groups, surrounded by photospheric plages and flocculi fields in
chromosphere. The area of spot is in several times less than the
total area of the chromospheric flocculi and coronal condensations
associated with this spot. The total area of solar spots in the
epoch of solar activity maximum occupies an area of 0,5 \% of the
hemisphere. The total area of the spots in solar-type stars can
reach 15 - 20 \%;

(c) The surface brightness of the active area increases with its
increasing in size and with increase in the number of spots in it.
On average, according to measurements at Skylab, (Schriver et al.
1985; Van Driel-Gesztelyi 2006) the surface brightness of active
regions is in 3 - 5 times higher than the brightness of the
chromospheric net, and a relative increase in luminance (contrast)
depends on the wave length;

(d) The variation of shortwave radiation are affected the existence
of "bright spots" - areas of the small bipolar region, in which
there are usually 2 - 3 magnetic loops, about 2500 km in diameter
and 12,000 km in length (Sheely \& Golub 1979). Comparison of  x-ray
photographs of the Sun, with simultaneous observations in the lines
of calcium K and $H_{\alpha}$ showed that "the bright x-ray points"
localized primarily on the borders of the cells of the chromospheric
net and in 83\% of cases coincide with prominent elements of the net
(Egamberdiev 1983). This testifies to the connection between "bright
x-ray points" and " bright points" of the net;

(e) Basically, the sources of x-ray radiation of the Sun (in
contrast to the ultraviolet radiation)  are situated in the coronal
loops, which are concentrated in active regions.  According to
observations of the satellite Skylab in the ultraviolet and x-ray
spectral bands all the coronal loops can be grouped into three main
groups, (Orral 1981):

the first group are the small arches, coming mainly across the
photospheric  section of the magnetic polarity, their temperature
is  about $2\cdot10^6 K$   , the concentration of electrons
$n_e\approx 3,5\cdot10^{9}cm^{-3}$ (in the surrounding space
$n_e\approx 2,5\cdot10^{9}cm^{-3}$);

the second group are the loops which connect  the magnetic poles out
of the line of section polarities, their  temperature of $10^5 <T<
10^6$ ;

the third group are the  cold loops  (chromospheric ribbons), which
are the bases of the first group, their temperature $T <5\cdot10^4
K$ (these loops are visible in $L_{\alpha}$ and other chromospheric
emission lines).

According to the results of observations of the satellite AE-E
Nimbus 7 the three-component model for the prediction of short-wave
radiation of the Sun was proposed (Lean et al. 1986). In this model
it is assumed that the flux of radiation is defined by three
components:

(I) is the constant component - the (BASAL) component, the sources
of which are evenly distributed on the solar disk, and do not change
in a cycle of activity;

(II) is evenly changing component - the radiation of the "active"
chromospheric net (which are also assumed to be uniformly
distributed over the disk, but its occurrence is considered to be
caused by the collapse of the Active Regions (AR) and, consequently,
its intensity should be proportional to their total areas;

(III) is the rapidly changing component - the intense radiation from
the plage areas, coinciding with the active regions.

In accordance with that the flux of radiation is calculated by the
formula (Lean et. al 1986):

$$  I = I_{\lambda Q} \Big\{ 1 + f_N \Big( C_{\lambda N}- 1 \Big) \Big\}
                      + 2 \pi F_ {\lambda Q}(1) \Sigma {A_i \mu_i R_{\lambda}(\mu_i)}
                      \Big ( C_{p \lambda} W_i -1 \Big ) \eqno (1) $$

where $I$ is the full flux of chromospheric emission, $I_{\lambda
Q}$ is the contribution of the
            constant component (BASAL), $C_{p \lambda}$ is the values of AR
            contrasts  and they are similar to contrasts from (Cook et all.,
            1980), $C_{N \lambda}$ is the value of "active network"
            contrast: they are equal to  $0.5 \cdot C_{p \lambda} $
            for continuum and $1/3 \cdot C_{p \lambda}$ for lines,
            $f_{N}$ is part of disk (without AR) that is occupied
            by the "active network".

            The second member in the right part of (1) describes
            emission from all AR on the disk;

            $A_i$ are values of their squares, $\mu_i$ describes the AR position:
            $\mu_i = {cos {\phi_i} cos {\theta_i}}$ (where $\phi_i$ and
            $\theta_i$ are the coordinates of AR number $i$).

            $R_\lambda(\mu_i)$ describes the relative change of the
            surface brightness $F_{\lambda Q}(\mu_i)$ with moving from
            center to edge of disk.

            The relative adding AR
            contribution to full flux from the different AR is
            determined by the factor $W_i$ that is linearly changed
            from the value $0.76$ to $1.6$ depending of the
            brightness ball of flocculus (according to ball flocculae changes from $1$ to $5$).

            So the "active network" part in all the surface without AR is determined by the AR decay,
            the next relationship between $f_N$ in time moment $t$
            and average values $A_i$ in earlier time is right:
            $$ f_N(t)= 13.3 \cdot 10^{-5} \cdot < \Sigma A_i(t-27)>  \eqno(2)$$
            where the time-averaging is taken for $7$ previous
            rotation periods, $A_i$ is measured in one million
            parts of the disk.

            Note that:

            first, the 27-day variation of the radiation caused by the movement
            and evolution of active regions;

            secondly, the 11-year cyclic changes in the
            significant contribution to make "active network".

\vskip12pt {\bf4. An empirical model of cyclic variations of the
values of fluxes of star's radiation
            in the $CaII$ lines of $H$ and $K$}
\vskip12pt

From the analysis of long-term variations of the solar streams in
the lines of ionized calcium $H$ and $K$, it follows that the total
fluxes of star's and Sun's radiation consist of three main
components: From the analysis of long-term variations of the solar
streams in the lines of ionized calcium $H$ and $K$, it follows that
the total fluxes of star's and Sun's radiation consist of three main
components:

(I) is the constant component, the so-called "BASAL" in solar
physics (below for the stars BASAL = $P_{min}$);

(II) is slowly changing in a cycle of activity background, including
the constant component $P_{min}$ ,  (for chromospheric  calcium
emission of the stars - $P_{CaII}(t)$ );

(III) - active regions on the disk (let's call this part of the flux
of chromospheric radiation of the stars - $S_{CaII}(t)$ ).

The total flux is equal to
$$ S_{CaII}(t) = P_{CaII}(t) +
S_{AR}(t)             \eqno(3)$$

Between the values of the fluxes of chromospheric radiation for the
Sun and the stars  $P_{CaII}(t)$ and $S_{CaII}(t)$, obviously, there
is a close connection that will be used in the future.

According to the data of observations (Lean et al. 1983; Lean 2005),
the maximum variation of the background of fluxes of solar radiation
in the radio band is about 20\%. This agrees well with the model of
radiation of the Sun in the line of hydrogen (Borovik et al. 1997),
where the maximum amplitude of the background radiation is also the
order of 20\%.

We introduce the factor of similarity k to forecast of cyclical
variations of chromospheric radiation of the stars. It equals to the
ratio of maximum amplitudes of the background radiation to the
maximum amplitude of the flux of star's radiation in a cycle of
activity:

$$k= (P_{CaII}^{max} - P_{min})/(S_{CaII}^{max} -
            P_{min})  \eqno (4)$$

\begin{figure}[h!]
 \centerline{\includegraphics{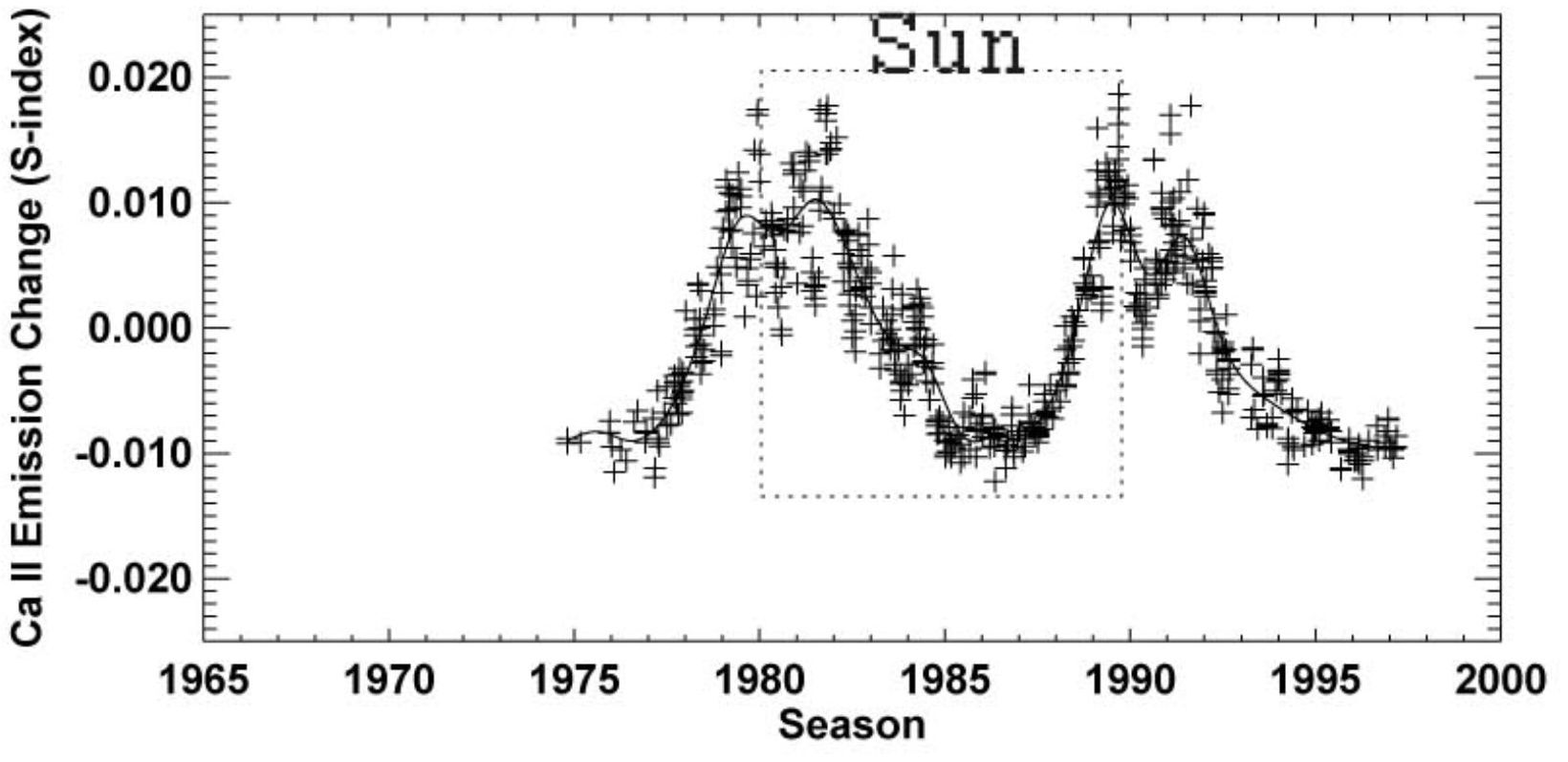}}
\caption{Variations of fluxes of  radiation of the Sun in
chromospheric lines, data set from (Radick et al. 1998)
}\label{Fi:Fig7}
\end{figure}

\begin{figure}[h!]
 \centerline{\includegraphics{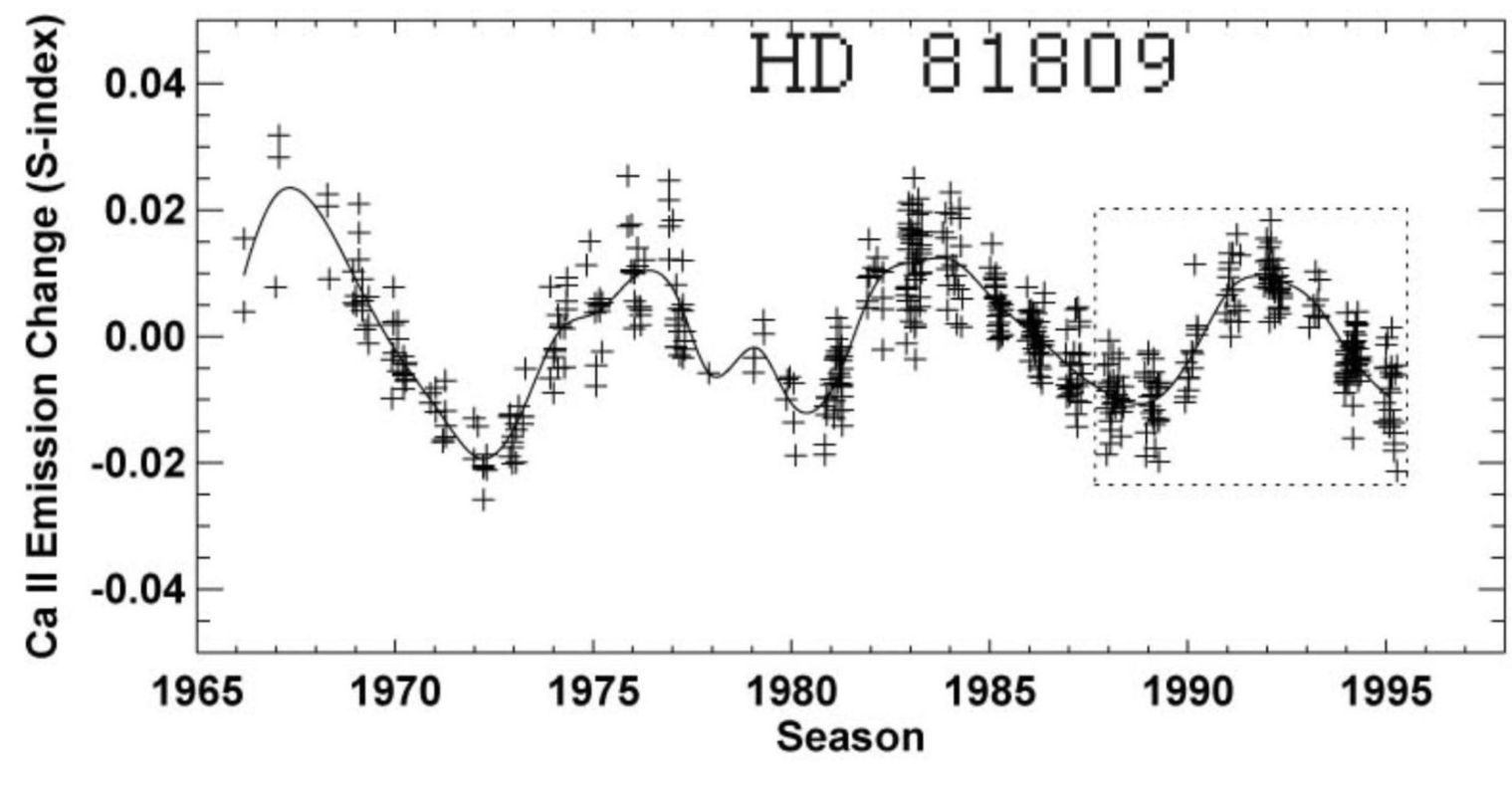}}
\caption{Variation of fluxes of the radiation of the HD 81809 star
 in chromospheric lines, data set from (Radick et al. 1998)
}\label{Fi:Fig8}
\end{figure}

In accordance with this, each value  $S_{CaII}(t)$ set in compliance
the value $P_{CaII}(t)= k \cdot S_{CaII}(t)$. Here the value
$S_{CaII}(t)$ was obtained from observation data and was presented
for the 13 stars with well-determined cycles of chromospheric
activity (star's group "EXCELLENT") and the Sun.

In Fig. 7 and Fig. 8 we present the smoothed curves of variations of
 the fluxes for the Sun and for one of the studied here stars - the star HD
81809, belonging to the group of "EXCELLENT" class.

From the observational data of chromospheric radiation (Radick et
al. 1998) and knowing the values of the periods of cycles
$T=T_{cyc}$ (see Table 2) for all 13 of the stars of the group of
"EXCELLENT" (see for example the data for the HD 81809 star in
Figure 8) we can define individual values of the flux of the
radiation $S_{CaII}(t/T)$ (with fairly good accuracy, determined by
the quality of observations).

Further, with a coefficient of similarity $k$  we can estimate these
values $P_{CaII}(t/T)$.

Then for each star we compute the coefficients $a$  and $b$ of the
regression equation (which will be further used for the forecasts of
the required moment of time in the future):

 $$ S_{CaII}(t)= a \cdot P_{CaII}(t) + b \eqno (5) $$

 Then when predicting the variations of the chromospheric fluxes for the 13 stars (Table 3),
 we apply the same approach as that used in the work (Bocharova \& Nusinov 1983)
 for the prediction of the solar background radiation in the "11-year" cycle.
 Modifying the formula of   (Bocharova \& Nusinov 1983), we obtain the following expression for the
 analytical approximation of the values for the background radiation of the stars (Bruevich 1999):

 $$ P_{CaII}(t)=P_{\min}\cdot \Big(1+\sin^4 \cdot{ {\pi \cdot t}
            \over{T}} \Big) \cdot e^{-{{\pi \cdot t}
            \over{T}} } \eqno (6) $$

 where is the minimum value of the background flow (BASAL),
 corresponding to the lowest value of the flux of radiation of the stars during the cycles
 of chromospheric activity and is determined from observations
 (on Fig. 9 and Fig. 10 the level of BASAL is shown with help of the solid line),
   $T$ - the period of the cycle chromospheric activity.

\begin{center}

\begin{table}
\caption{Observed parameters of 13 "EXCELLENT" class stars and
regression coefficients $a$ and $b$ calculated from (5)}
\bigskip
 \begin{tabular}{|l|c|c|c|c|c|c|c|}
  \hline

  Object & B - V & $T_{cyc}$,  & $P_{\min}$ & a & b & $ {\Delta S_{CaII}^{\max}}/$& ${\Delta S_{AR}^{\max}}/$  \\
         &       &   year      &            &   &   &  ${P_{\min}}$           &  $P_{\min},\%$             \\ \hline
  Sun       & 0.66 & 10   & 0.162 & 1.19 & -0.031 & 23.4 & 3.4  \\ \hline
  HD 81809  & 0.64 & 8.2  & 0.155 & 1.13 & -0.020 & 22.6 & 2.6  \\ \hline
  HD 152391 & 0.76 & 10.9 & 0.32  & 1.56 & -0.180 & 31.6 & 11.3 \\ \hline
  HD 103095 & 0.75 & 7.3  & 0.17  & 1.23 & -0.040 & 24.7 & 4.7  \\ \hline
  HD 184144 & 0.80 & 7.0  & 0.19  & 1.45 & -0.085 & 28.9 & 8.9  \\ \hline
  HD 26965  & 0.82 & 10.1 & 0.18  & 1.39 & -0.07  & 27.8 & 7.8  \\ \hline
  HD 10476  & 0.84 & 9.4  & 0.17  & 1.61 & -0.104 & 32.4 & 12.3 \\ \hline
  HD 166620 & 0.87 & 15.8 & 0.175 & 1.43 & -0.075 & 28.6 & 8.6  \\ \hline
  HD 160346 & 0.96 & 7.0  & 0.24  & 1.88 & -0.21  & 37.5 & 17.5 \\ \hline
  HD 4628   & 0.88 & 8.4  & 0.19  & 1.96 & -0.183 & 39.4 & 19.4 \\ \hline
  HD 16160  & 0.98 & 13.2 & 0.19  & 1.61 & -0.116 & 32.6 & 12.6 \\ \hline
  HD 219834B& 0.91 & 10.0 & 0.17  & 1.92 & -0.157 & 38.2 & 18.2 \\ \hline
  HD 201091 & 1.18 & 7.3  & 0.51  & 1.85 & -0.434 & 37.2 & 17.2 \\ \hline
  HD 32147  & 1.06 & 11.1 & 0.22  & 1.67 & -0.147 & 45.4 & 25.4 \\ \hline

 \end{tabular}
 \end{table}
\end{center}

For the stars $T=T_{cyc}$ can vary from 6 to 23 years, according to
data(Baliunas et al. 1995). However, for stars of the group
"EXCELLENT" this value  varies in a narrow framework and is closed
to the solar period in 11 years. The variable $t$ is the current
time from the beginning of the cycle, expressed as a fraction of
cycle (for example,$t = 0,1\cdot T, 0,2\cdot T... 0,9\cdot T, T$).

Simulated flux $ S_{CaII}(t)$ for solar-type stars we can calculate
from the regression equation (5), using values $a$  and $b$ , as
well as having calculated in advance the value $ P_{CaII}(t)$ with
help of the formula (6).

In the Table 3 we show the results of our calculation of the
coefficients  $a$   and $b$   for 13 stars, having a well-determined
long-term cyclical variability of their fluxes  $ S_{CaII}$.

Also we presented the values of relative maximum variation of fluxes
in a cycle of activity, as a total fluxes $ {\Delta S_{CaII}^{max}}
            / {P_{min}}$ , as well as of that part
of the fluxes, which is responsible only for the contribution to the
radiation from active regions on the disk $ {\Delta S_{AR}^{max}}
            / {P_{min}}$ .

In Table 3 we present also the relative
            full flux variation in activity cycle maximum: ($ {\Delta S_{CaII}^{max}}
            / {P_{min}}$) and relative AR adding flux in activity cycle maximum: ($
            \Delta S_{AR}^{max} / P_{min}$). The value $P_{mim}$ -
            that is equal to BASAL emission for different we can determine from
            (Baliunas et al/ 1995) data, see Fig. 9 and Fig. 10.

\begin{figure}[h!]
 \centerline{\includegraphics{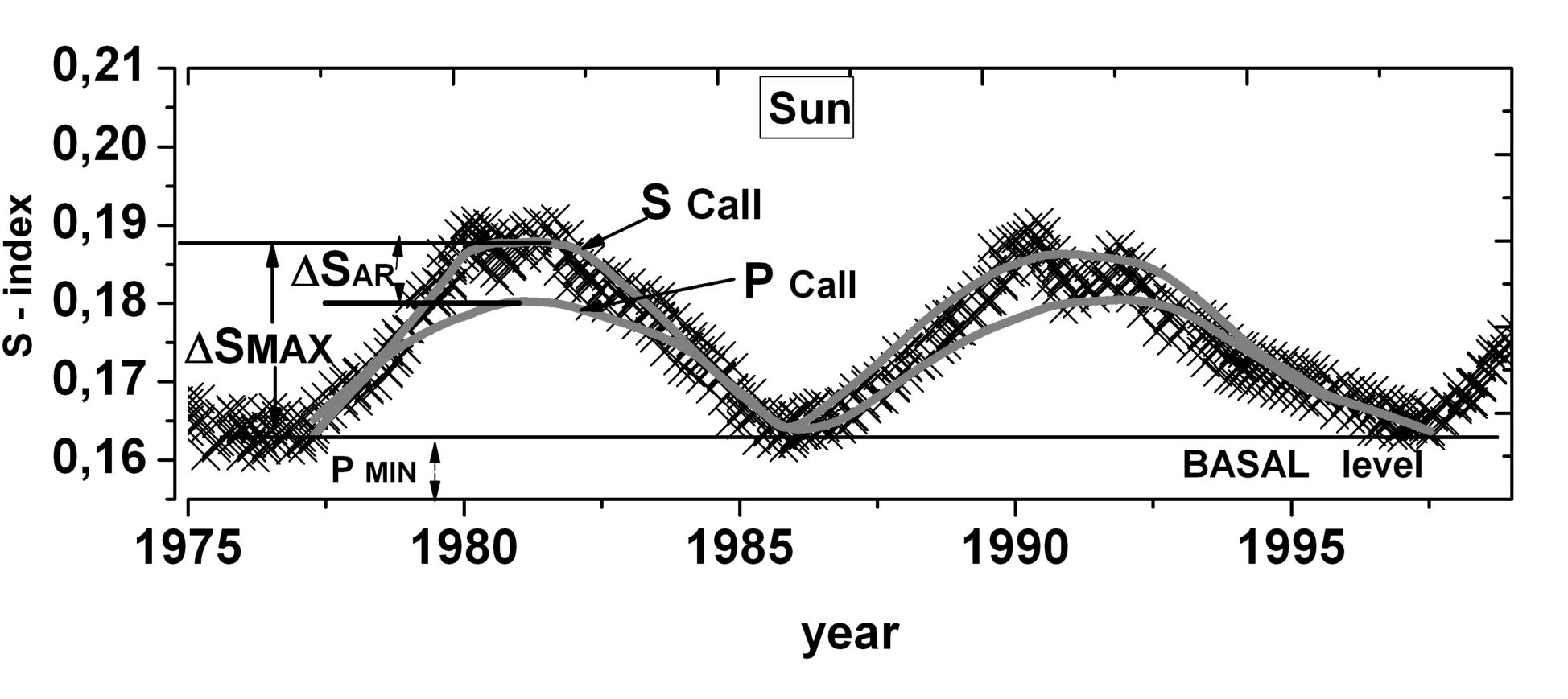}}
\caption{Variations of  fluxes of radiation of the Sun in
chromospheric lines according to observations the Observatory Mount
Wilson (Lockwood et al. 1997) - crosses and calculated according to
the formulas (5) and (6) the model curves for $S_{CaII}$  and
$P_{CaII}$ } \label{Fi:Fig9}
\end{figure}

\begin{figure}[h!]
 \centerline{\includegraphics{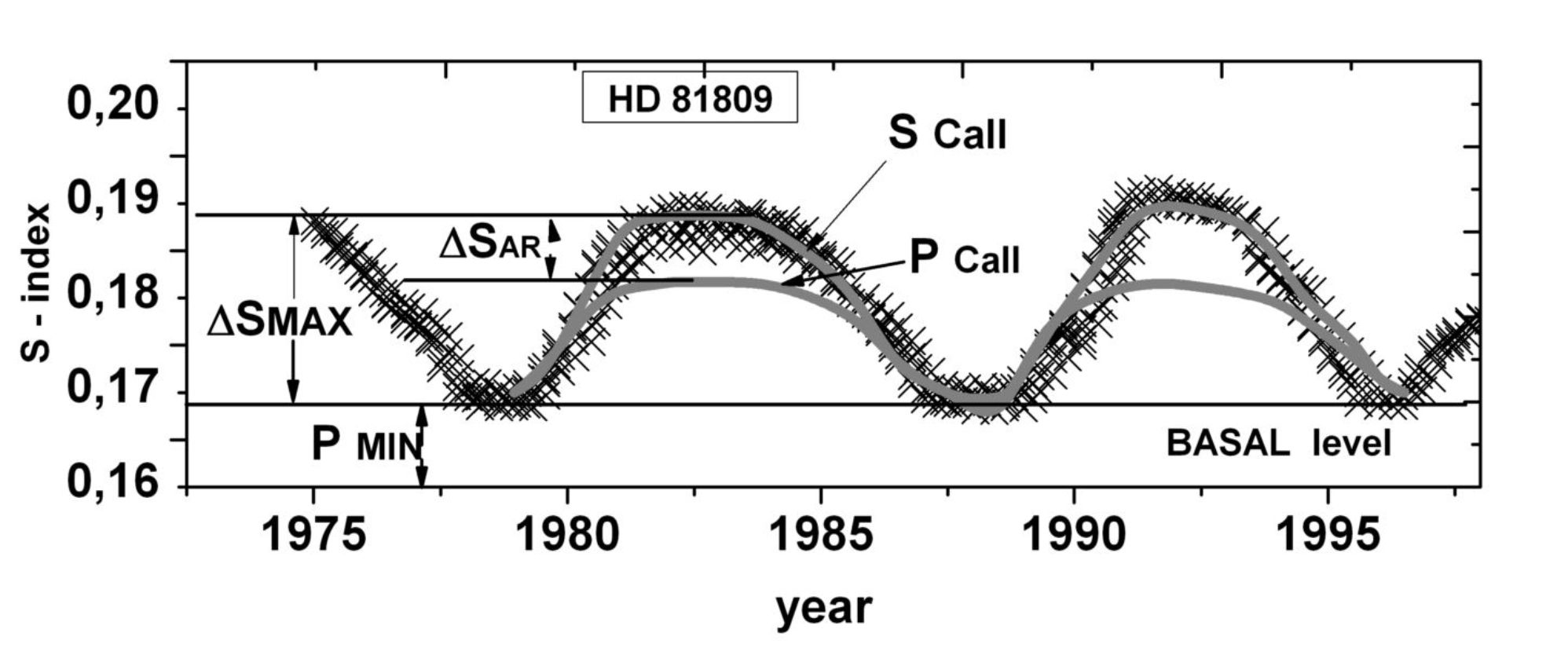}}
\caption{Variations of  fluxes of radiation of the HD 81809 star in
chromospheric lines according to observations the Observatory Mount
Wilson (Lockwood et al. 1997) - crosses and calculated according to
the formulas (5) and (6) the model curves for $S_{CaII}$ and
$P_{CaII}$} \label{Fi:Fig10}
\end{figure}

In Table 3 and at Fig. 9 and Fig. 10 we use the following notations:

 $P_{\min}$  - the constant component of the radiation of the stars - (BASAL),
 $ P_{CaII}$  - the background radiation, changing in a cycle of activity,
 $ S_{CaII}$ - the complete flux of radiation ,
  ${\Delta S_{CaII}^{\max}}$ - the maximum amplitude of the variations of radiation in a cycle of activity
  ${\Delta S_{AR}^{\max}}$- the contribution to the total flux of the active regions.

Thus, the procedure of the forecast of model curves of cyclic
variations of fluxes of chromospheric  radiation of stars in the
11-year time scale consists of the following:

1. For the prediction of the radiation intensity of use data from
Table 2 (take the value  $P_{\min}$ for the selected stars from
column 5). In Fig. 9 and Fig. 10 - the level of BASAL=$P_{\min}$.

2. From Fig. 10 (star HD 81809) and from (Lockwood et al. 2007) for
the stars of the group "EXCELLENT" we define the moment of the
beginning of the chromospheric cycle and $t/T$ for the selected us
time of the forecasting flux.

3. Then using equation (6) calculate the value $ P_{CaII}(t)$  for
the moment $t$.

4. Using equation (5) we calculate the forecasting flux  $
S_{CaII}(t)$ for the moment $t$.

\bigskip

{\bf5. Conclusions}

\bigskip

The large amount of data of regular observation of solar radiation
in different spectral ranges does preferred for us the conduct of
wavelet-analysis to study the cycles of magnetic activity of the
Sun.

The wavelet-analysis helped us to see a set of values of periods of
cycles besides 11-year cycle:

-   long duration cycles at 22-year, 40-50 year and 100-120 year
time scales;

-   short duration cycles at quasi-biennial and 1,3-year time
scales.

Unfortunately in case of stars we have time series of  observations
in one - two spectral ranges which are less informative and short in
time. This often forcing us to be limited to Fourier analysis in the
study of cyclic recurrence of radiation from stars.

When we analyze results of our predictions (in Table 3 we presented
the observed values
      that we're discussed in this issue and our estimations as $ {\Delta S_{CaII}^{max}}
            / {P_{min}}$ and $\Delta S_{AR}^{max} / P_{min}$) some conclusions can be
      made:

- we can see from the last column of Table 3 that, among the stars
of the group "EXCELLENT" of spectral class K (with the best quality
of cyclicity) the active regions may provide the excess above the
background flux (consisting of constant for each star
BASAL-radiation and of radiation from changing the brightness during
the cycle of chromospheric network) in the maximum of the cycle of
up to 10-20\%;

- the most bright flocculi, which are in 2 times brighter than the
unperturbed of chromosphere (cell centers of chromospheric network)
should occupy the area of 5 - 10\% of the surface of the star;

- in case of the young stars of spectral classes K and M without
cycles this value (area that bright flocculi should occupy) is much
larger (up to 50\%), and for the stars of spectral type G with the
well determined cycles of activity, much less (1-3 \%). For the Sun
this value is only 0,1 \%.

\bigskip

{\bf References}

\bigskip

1. Baliunas, S.L., Donahue, R.A., et al. (1995) Astrophys.
        J., {\bf 438}, 269.

2. Beer J., (2000), Space Sci. Rev., {\bf 94}, 53.

3. Borovik, V.N., Livshitz, M.A., Medar, V.G. (1997)
        Astronomy Reports, {\bf 41}, N6, 836.

4. Bruevich, E.A., Kononovich E.V. (2011) Moscow University Physics
Bull., {\bf66}, N1, 72; ArXiv e-prints, (arXiv:1102.3976v1)

5. Bruevich, E.A., Ivanov-Kholodnyj G.S. (2011) ArXiv e-prints,
    (arXiv:1108.5432v1).

6. Bruevich, E.A. (1999) Mosc. Univ. Phys. Bull., Ser. 3, {\bf No6},
48.

7. Cook, J.W., Brueckner, G.E., Van Hoosier,
        M.E., (1980) J.Geophys.Res., {\bf A85}, N5, 2257.

8. Van Driel-Gesztelyi, L. (2006) in: V. Bothmer, A.A. Hady(eds.),
Solar Activity and its Magnetic Origin, IAUS 233, 205.

9. Egamberdiev, S. A. 1983, Soviet Astronomy Letters, {\bf9}, 385.

10. Ivanov-Kholodnyj, G.S. \& Chertoprud, V.E., (2008)
  Sol.-Zemn. Fiz., {\bf 2}, 291.

11. Khramova, M.N., Kononovich, E.V. \& Krasotkin, S.A., (2002)
Astron. Vestn., {\bf 36}, 548.

12. Kollath, Z., Olah, K. (2009) Astron. and Astrophys,  {\bf 501},
695.

13. Lean J.L., Scumanich A., (1983) J. Geophys. Res., {\bf
        A88}, N7, 5751.

14. Lishits, I.M., Obridko, V.N., (2006) Astronomy Reports, {\bf
83}, N 11, 1031.

15. Lockwood, G.W., Skif, B.A., Radick R.R., Baliunas, S.L.,
    Donahue, R.A. and Soon W., (2007) Astrophysical Journal Suppl., {\bf 171}, 260.

16. Noyes, R. W.; Hartmann, L. W.; Baliunas, S. L.; Duncan, D. K.;
Vaughan, A. H., (1984) Astrophys. J., {\bf 279}, 763.

17. National Geophysical Data Center Solar and Terrestrial Physics,

http://www.ngdc.noaa.gov/stp/solar/sgd.html

18. Orral, F.Q. (1981) Space Sci. Rev., {\bf28}, N4, 423.

19. Radick, R.R., Lockwood, G.W., Skiff, B.A., Baliunas, S.L. (1998)
  Astrophys. J. Suppl. Ser., {\bf 118}, 239.

20. Rivin,~Yu. R., (1989) The cycles of The Earth and the Sun,
Moscow, Nauka.

21. Schriver, C.J., Zwaan, C., Maxon, C.W., and Noyes,
        R.W., (1985). Astron. and Astrophys., {\bf 149}, N1, 123.

22. Sheely, N.R., Jr. Golub, L., (1979) Solar Phys., {\bf63}, N1,
119.

23. Vernazza, J.E., Avrett, E.H., Loeser, R., (1981) Astrophys.
        J.Suppl.Ser., {\bf 45}, N4, 635.

24. Vitinsky,~Yu.I., Kopecky,~M., Kuklin,~G.B. (1986) The statistics
of the spot generating activity of the Sun, Moscow, Nauka.

25. Wilson, O.C., (1978). Astrophysical J., {\bf 226}, 379.

\end{document}